# Exploiting Machine Learning and Disequilibrium in Galaxy Clusters to Obtain a Mass Profile

Mark J. Henriksen  and Prajwal Panda
University of Maryland, Baltimore County Physics Department, 1000 Hilltop Circle, Baltimore, MD 21250, USA

## Abstract

We use 3D $k$-means clustering to characterize galaxy substructure in the A2146 cluster of galaxies ($z=0.2343$). This method objectively characterizes the cluster's substructure using projected position and velocity data for 67 galaxies within a 2.305 Mpc circular region centered on the cluster's optical center. The optimal number of substructures is found to be four. Four distinct substructures with rms velocity typical of galaxy groups or low-mass subclusters, when compared to cosmological simulations of galaxy cluster formation, suggest that A2146 is in the early stages of formation. We utilize this disequilibrium, which is so prevalent in galaxy clusters at all redshifts, to construct a radial mass distribution. Substructures are bound but not virialized. This method is in contrast to previous kinematical analyses, which have assumed virialization, and ignored the ubiquitous clumping of galaxies. The best-fitting radial mass profile is much less centrally concentrated than the well-known Navarro–Frenk–White profile, indicating that the dark-matter-dominated mass distribution is flatter pre-equilibrium, becoming more centrally peaked in equilibrium through the merging of the substructure.

*Unified Astronomy Thesaurus concepts:* Galaxy clusters (584); Astroinformatics (78); Large-scale structure of the universe (902); Dark matter distribution (356)

## 1. Introduction

Early X-ray observations by the Einstein observatory showed a nonequilibrium, bimodal substructure (Forman et al. 1981; Henry et al. 1981), suggesting that structure formation proceeds via hierarchical merger. The Rosat mission, with an increase in sensitivity of four over Einstein, showed that substructure is common in galaxy clusters (Scheucker et al. 2001). Evidence of multiple mergers predating the bimodal morphology, for example in A98, originally thought to be bimodal, is present in the Chandra observations (Sarkar et al. 2023). Cold dark matter (CDM), in which baryonic matter follows the collapse of a dark matter distribution consisting of particles that have low thermal velocities (Peebles 1983), features hierarchical structure formation. Simulations of cluster formation in a cosmological setting provide a general sequence of cluster formation in which smaller structures coalesce into a major merger stage involving two nearly equal subclusters, which then merge into a massive cluster that continues accreting groups of galaxies (Wu et al. 2013). Modern X-ray observations provide evidence that substructure is related to mergers through the detection of shock-heated gas. Twisted core X-ray isophotes, seen in many clusters, match the simulations suggesting that clusters form and evolve not only through mergers but also through interactions (Tittley & Henriksen et al. 2005). The prevalence of substructure in galaxy clusters has resulted in galaxy kinematic data falling out of favor as a probe of cluster masses because modelers typically assume virialization of the galaxy distribution.

In this Letter, we investigate tools to objectively characterize the substructure in the A2146 galaxy cluster. We then use the kinematics of the substructure to derive the mass distribution of this galaxy cluster, which is clearly not in equilibrium.

## 2. Data

Galaxies with position and velocity in the NASA/IPAC Extragalactic Database are extracted within a circle of radius 10′ centered on the X-ray center (239.03708, 66.3558; Piffaretti et al. 2011). This is a radius of 2.305 Mpc corrected for the cosmology defined by: $H_o = 67.8$, $\Omega_m = 0.308$, $\Omega_{vac} = 0.692$, and $z = 0.2343$. Sixty-seven galaxies meet the criteria. One galaxy was eliminated using a clipping algorithm. The mean velocity from the redshift data is 69,736 km s$^{-1}$ and the standard deviation is 1334 km s$^{-1}$. The standard deviation is large but typical of a cluster in a state of disequilibrium.

## 3. Analysis

We applied the $k$-means clustering method to the 3D kinematic data set in order to characterize the substructure. $k$-means belongs to the category of hard clustering algorithms in which each data point is assigned to exactly one cluster. $k$-means assigns each point to the cluster whose centroid has the smallest distance to that data point. The algorithm recalculates the centroids of the $k$ clusters based on newly assigned data points. This process continues iteratively until the cluster centroids remain stable. One of the limitations of $k$-means is sensitivity to outliers. Outliers can significantly affect the position of cluster centroids and may lead to poor clustering results. This problem has been mitigated in our galaxy kinematic data set by clipping the data. The number of clusters is input for each application of $k$-means. Runs ranging from three to seven clusters are analyzed.

To determine the optimal number of clusters for A2146, we use the silhouette and the elbow methods. The silhouette method balances the cohesiveness of a cluster with the separation between clusters. A score related to this balance is calculated by measuring how similar each data point is to its own cluster compared to other clusters. The number of clusters is then varied and the silhouette score is recalculated. A higher silhouette score suggests that the data point is well-matched to its own cluster. The optimal number of clusters is obtained from the highest score. The high score occurs when there are four clusters (see Figure 1). The elbow method evaluates the trade-off between the number of clusters and the total within-cluster variation. The sum of the squared distance between data points and their assigned cluster centroid, the within-cluster sum of squares (WCSS), is calculated for the range of cluster numbers. The optimal number of clusters is the point at which the graph forms an elbow since the reduction in within-cluster variation begins to plateau. This occurs when there are four clusters (see Figure 2).

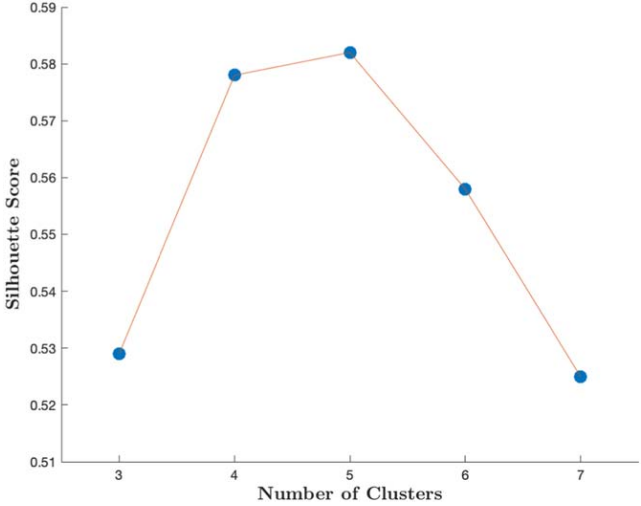

Figure 1. Four clusters is the optimum partitioning for the silhouette method.

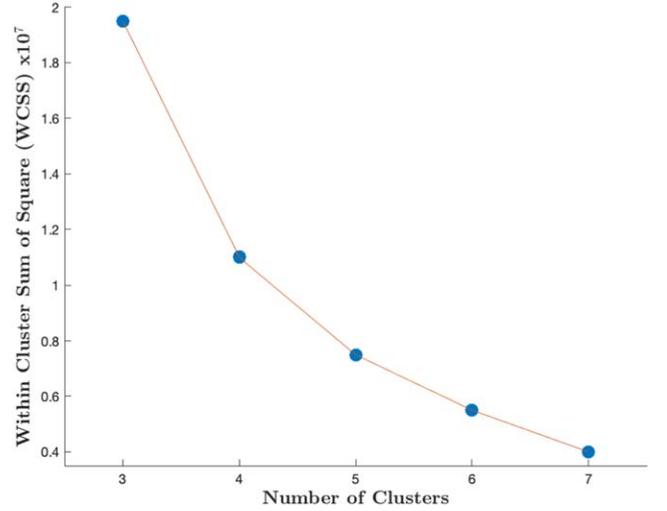

Figure 2. Four clusters is the optimum partitioning for the elbow method.

## 4. Results

The silhouette and elbow methods both favor four clusters in the $k$-means analysis. We ran the clustering algorithm fixed at the optimum number, four, four times. Four runs are utilized because the $k$-means algorithm begins with four randomly chosen regions. For each run, the algorithm chooses one of the regions as the initial starting point and then begins the clustering process. Thus, there are four unique runs; on the fifth run, all clusters are duplicates of those found in the previous runs. Within the four runs, 2 of the 16 clusters had duplicates, with one of the clusters having two duplicates. Of the 16 clusters, 13 are unique. Using 13 clusters as the data set includes the inherent uncertainty of the clustering algorithm since it reflects variations due to the initial conditions. That is, rather than converging on 4 clusters that are found in each subsequent run, there are 13 unique clusters found over the 4 runs. The mean velocity and rms velocity is calculated for each of the clusters (see Table 1). We then made a sanity check that the results are physically reasonable based on our general knowledge of galaxy cluster substructure. An ongoing merger within a galaxy cluster has both spatial substructure and velocity substructure in the kinematic data. The velocity substructure may have multiple peaks or have a very large dispersion, as we find here. The assumption is that the substructures are self-gravitating units. Thus, the rms velocity of the substructure should be typical of a group or subcluster.

Using the 3D $k$-means, we find that the 13 clusters have rms velocity ranging from 284 to 566 km s$^{-1}$. These are well within the range of galaxy groups (Xue & Wu 2000). The centroid of each cluster is given in Table 1. The centroid of the cluster of galaxies is (239.0355, 66.3598), which is the mean R.A. and decl. in degrees of the 66 galaxies. Table 1 also gives the distance from the center of A2146 in arcseconds and in kiloparsecs using the cosmology-corrected scale of 3.842 kpc arcsecond$^{-1}$. Figure 3 shows a 3D scatterplot of one of the four runs that is representative of $k$-means clustering. All four clusters are fairly close to the center of the cluster. This may have implications for the evolutionary stage of the cluster. Figure 4 shows the mean cluster velocity versus distance for the 13 clusters. We use these data and the rms velocities to further interpret the evolutionary state of the cluster and its mass distribution.

## 5. Galaxy Cluster Radial Mass Profile Determination During Disequilibrium

Galaxy cluster masses have been determined in three primary ways: virial analysis of galaxy kinematic data, hydrostatic analysis of the X-ray-emitting intracluster medium, and gravitational lensing. The fact that clusters continue to experience mergers has marginalized dynamical mass determinations based on both gas and galaxy dynamics because they have relied on the assumption of equilibrium. Here we propose a way to utilize the ubiquity of cluster mergers in a new approach of mass determination. Multiple substructures appear in cosmological simulations of cluster formation at two early stages (i.e., before the bimodal subclusters that then merge to form the single component cluster). The first stage where multiple substructures are seen is during the initial spherical collapse. We assume that each substructure starts at rest on the outside edge of the collapsing sphere of mass. We use a simple model of spherical collapse to predict the infall velocities and compare them to those measured from $k$-means clustering.

We begin with the differential equation for spherical collapse,

$$\frac{d^2r}{dt^2} = \frac{-GM_r}{r^2}. \quad (1)$$

Equation (1) is integrated to obtain the infall velocity, $v$. The initial density, $\rho_0$, depends on the initial radius, $r_0$, and the spherical mass, $M$.

Table 1
Optimum k-means Clustering Parameters

| R.A. (deg) | Decl. (deg) | Radius (arcsec) | Radius (kpc) | Mean Vel. (km s$^{-1}$) | rms Vel. (km s$^{-1}$) | Infall Vel. (km s$^{-1}$) | No. Gal. |
|---|---|---|---|---|---|---|---|
| 239.0367 | 66.3598 | 4.95 | 19 | 70,551 | 397 | 815 | 17 |
| 239.0376 | 66.3582 | 6.51 | 25 | 70,415 | 458 | 679 | 20 |
| 239.0403 | 66.3578 | 10.0 | 38 | 70,428 | 450 | 692 | 21 |
| 239.0404 | 66.3578 | 10.09 | 39 | 70,153 | 310 | 417 | 18 |
| 239.0331 | 66.3567 | 11.69 | 45 | 67,941 | 500 | 1822 | 13 |
| 239.0367 | 66.3629 | 11.98 | 46 | 69,309 | 355 | 427 | 26 |
| 239.0353 | 66.3559 | 14.04 | 54 | 67,893 | 490 | 1843 | 12 |
| 239.0453 | 66.3573 | 16.77 | 64 | 71,639 | 566 | 1903 | 15 |
| 239.0299 | 66.3646 | 18.32 | 73 | 67,990 | 514 | 1746 | 14 |
| 239.0368 | 66.3540 | 20.96 | 80 | 67,979 | 502 | 1757 | 13 |
| 239.0264 | 66.3650 | 22.87 | 88 | 69,176 | 287 | 560 | 24 |
| 239.0493 | 66.3564 | 23.38 | 90 | 71,907 | 506 | 2171 | 10 |
| 239.0248 | 66.3655 | 25.68 | 99 | 69,104 | 284 | 632 | 22 |

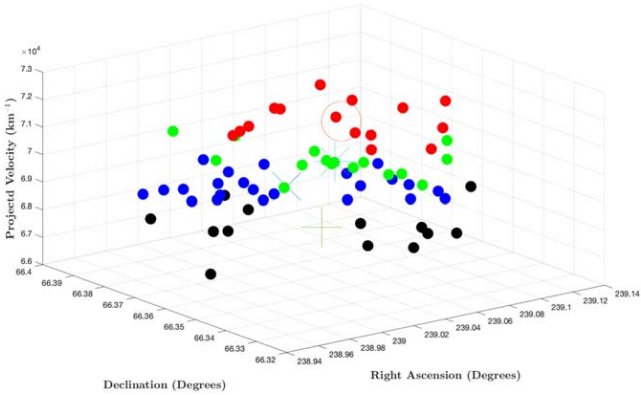

**Figure 3.** A graphical example of a single run with four substructures. Symbols mark the substructure centers.

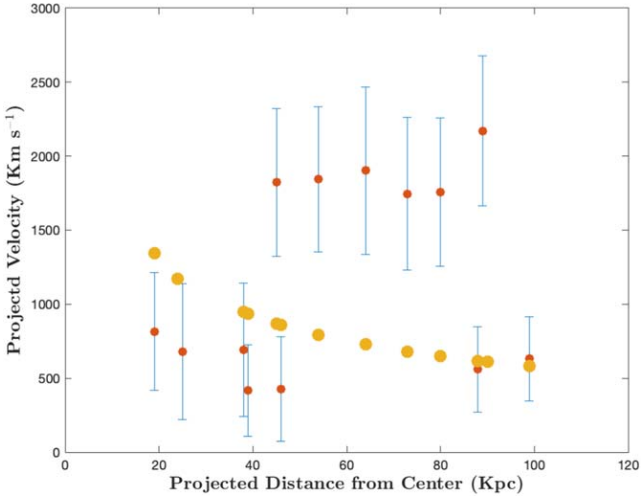

**Figure 4.** Infall velocity vs. radius compared to a spherical infall model.

$$\frac{dr}{dt} = -\left[\frac{8\pi}{3}G\rho_0 r_0^2\left(\frac{r_0}{r} - 1\right)\right]^{1/2} \quad (2)$$

$$v = 523\frac{\text{km}}{\text{s}}\left(\frac{\rho_0}{\rho_c}\right)^{1/2}\left(\frac{R_0}{8\text{ Mpc}}\right)\left(\frac{8}{r_{\text{mpc}}} - 1\right)^{1/2}. \quad (3)$$

The initial cluster density is scaled as a fraction of the critical density, $8.5 \times 10^{-30}$ g cm$^{-3}$, and is equal to $\Omega_m$. The observed infall velocity is the magnitude of the difference between the mean velocity of the substructure, given in Table 1, and the mean cluster velocity, 69,736 km s$^{-1}$. The velocity at any radius is a function of the matter density, $\Omega_m$ only, for a nominal region of 8 Mpc in Equation (3). At the smallest radius, 19 kpc, two nominal matter densities, 0.3 and 0.04, both give a very large infall velocity (5871 and 2144 km s$^{-1}$, respectively) compared to observations. We reduce the size of the initial protocluster to 3 Mpc in order to more easily show the shape of the radial velocity profile for comparison to observations (see Figure 4).

The general trend in velocities for spherical infall, higher velocities at smaller radii, is not consistent with the general observational trend, lower velocities at smaller radii. Lower velocities in the center may be due to dynamical friction. We conclude that the structures have gone through a first pass and experienced a degree of kinetic energy dissipation. This interpretation is supported by the fact that the subclusters appear to overlap, spatially.

We now assume only that the subclusters are bound, but not in equilibrium, to obtain a mass profile. In order to derive a radial mass profile from the radial velocity data, we first fit a power law to the $(r, v)$ cluster data,

$$v(r) = A \times r^b. \quad (4)$$

The error bars used in the fit are the standard deviation of the mean velocity for each cluster and appears as a weighting factor in the fitting procedure. The units of $v$ are km s$^{-1}$ and $r$ is in kiloparsecs. The best-fit parameters, with 95% confidence uncertainties, are $A = 276 -485/+1037$ and $b = 0.41 -0.23/1.05$. Though the confidence limits for both parameters are very large, it is apparent that with very high confidence, the velocity increases with radius. The best-fit $v(r)$ profile is shown with the data in Figure 5.

A radial mass profile, $M(r)$, is calculated by substituting the best-fit $v(r)$ into the equation of energy balance based on the condition that each substructure is bound. That allows the calculation of the mass interior to any radius. The resulting radial dependence is given in Equation (5),

$$M(r) \sim r^{1.82}. \quad (5)$$

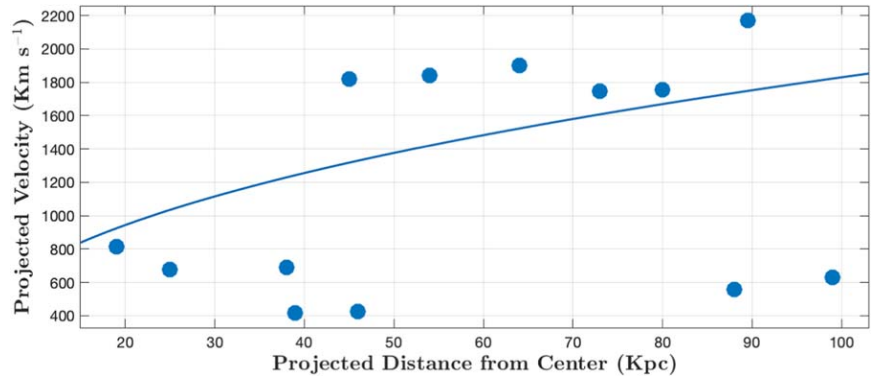

**Figure 5.** Best-fit power law to infall vs. distance data favors 0.41, consistent at a 96% confidence interval.

The radial profile of the cumulative mass is shown in Figure 6. For comparison, the Navarro–Frenk–White (NFW) profile (Navarro et al. 1996; Bertone 2010) is also shown in Figure 6, using Equation (6) and normalizing at 1.28 Mpc. The shapes are different in the center. We further investigate this apparent difference by fitting the cumulative NFW mass profile to the observed cumulative mass profile. The scaling parameter $r_s$ and the normalization, $A$, are the free parameters in Equation (6),

$$M(<r) = A \times \left[\ln\left(\frac{r_s + r}{r_s}\right) - \left(\frac{r}{r_s + r}\right)\right]. \quad (6)$$

The best-fit ($A$, $r_s$) is (21046, 5822). The 95% confidence range is $16{,}000 < A < 26{,}000$ and $5048 < r_s < 6596$. $A$ is $4\pi\rho_0(r_s)^3$. The best-fit parameters give a central density of $6.3 \times 10^{-27}$ g cm$^{-3}$. Figures 7 and 8 show that the NFW profile can be forced to fit the observed mass profile, though the magnitude of the best-fit parameters is extremely large.

## 6. Discussion

Utilizing the dynamical disequilibrium in clusters to characterize substructure in order to obtain the mass distribution has the potential to provide an interesting constraint on cosmological parameters that use cluster mass, the dark matter distribution on cluster scales, and cluster formation physics. The $k$-means clustering method provides reasonable substructure information, based on the number of galaxies and rms velocity. The fact that all galaxies have to be in a substructure may be too narrow of a constraint and may not be a realistic requirement. This requirement should raise the rms velocity of a substructure, yet the values obtained are all physically reasonable. Applying two methods to evaluate the optimum number of substructures clearly favors four in both cases for A2146. A multiple-substructure morphology is also apparent in cluster simulations as an early stage in the hierarchical cluster formation process. This morphology appears in simulations during the initial collapse and again after a pass that precedes the next merger product, a bimodal morphology. A spherical infall model predicts a radial velocity dependence that peaks in the center and decreases with increasing radius. The gravitationally bound model predicts a velocity profile that increases with radius and more accurately reflects the radial infall velocity profile for A2146. The best-fit radial velocity model results in an integrated mass profile that is quite large, approaching $3.6 \times 10^{15}$ solar masses at 1.28 Mpc, though this may be due to the substructures partially reflecting the spherical collapse stage. The derived radial mass profile can only match an NFW profile for a very large scaling radius. This large scaling radius acts to flatten out the radial mass profile of the NFW, known for its steep central profile, indicating that the NFW profile is not an accurate physical model for a cluster in the early stages of formation. The concentration factor, $c$, in the NFW density profile ranges from ∼5 to 10 on cluster mass scales (Navarro et al. 1996). The relationship between $r_{200}$ and $r_s$ predicts a value of ∼150–300 kpc for the latter parameter, estimating $r_{200}$ to be 1.5 mpc. The best-fit $r_s$ obtained from fitting the observed mass profile for A2146 is 20 times larger, supporting the atypical values found by force-fitting an NFW profile. Figure 6 shows a comparison of the integrated mass profile for A2146 compared to the NFW, normalized at 1.28 Mpc using a nominal scaling radius of 250 kpc, typical of isothermal beta-model fits to the final stage clusters and in the range of typical scaling radii. The NFW profile is more centrally peaked. During the later stages of cluster formation, significant concentrating of mass occurs. This is presumably accomplished through hierarchical structure mergers and the dissipation of gas via radiative cooling, both of which serve to steepen the radial profile in the inner regions of mass halos (Gnedin et al. 2004). This agrees with what we have found here; that a flatter mass distribution precedes a final, more centrally peaked profile.

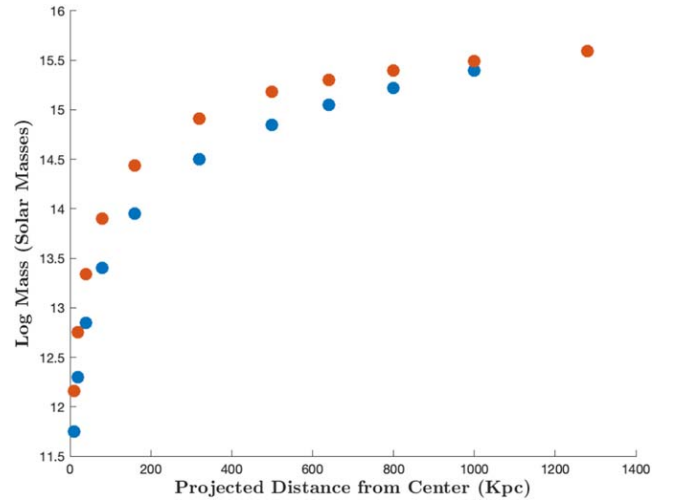

**Figure 6.** Derived cumulative mass profile (blue) with NFW shown for comparison.

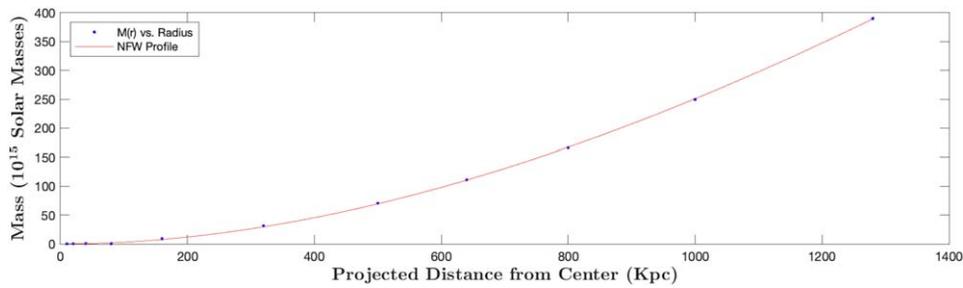

**Figure 7.** Best fit of NFW mass profile to the mass profile derived from the data.

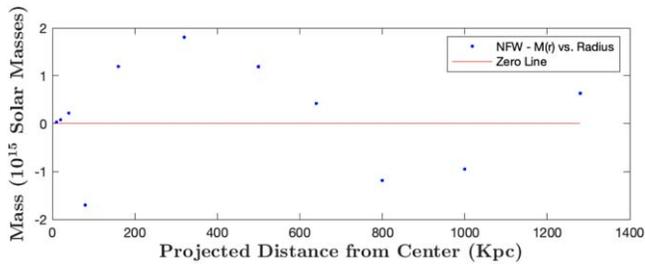

**Figure 8.** Residuals for best-fit NFW.

## ORCID iDs

Mark J. Henriksen  https://orcid.org/0000-0003-0530-8736